\documentclass[prd,aps,showpacs,tightenlines,nofootinbib,preprint,
preprintnumbers,eqsecnum]{revtex4}
\usepackage{graphicx,color,amsmath,amsxtra}
\usepackage{amssymb}

\newcommand{\bear}{\begin{array}}  \newcommand{\eear}{\end{array}}
\newcommand{\bea}{\begin{eqnarray}}  \newcommand{\eea}{\end{eqnarray}}
\newcommand{\beq}{\begin{equation}}  \newcommand{\eeq}{\end{equation}}
\newcommand{\bef}{\begin{figure}}  \newcommand{\eef}{\end{figure}}
\newcommand{\bec}{\begin{center}}  \newcommand{\eec}{\end{center}}

\newcommand{\Eqn}[1]{&\hspace{-0.2em}#1\hspace{-0.2em}&}

\def\Vec#1{\mbox{\boldmath $#1$}}
\def\Vecs#1{\mbox{\boldmath\tiny $#1$}}
\def\Lap{{\mathop{\Delta}\limits^{(3)}}}

\def\be{\begin{equation}}
\def\ee{\end{equation}}
\def\bea{\begin{eqnarray}}
\def\eea{\end{eqnarray}}
\def\beq{\begin{eqnarray}}
\def\eeq{\end{eqnarray}}

\begin{document}

\title{
Inflation and late-time cosmic acceleration \\
in non-minimal Maxwell-$F(R)$ gravity \\
and the generation of large-scale magnetic fields
}

\author{Kazuharu Bamba$^1$ and Sergei D. Odintsov$^2$\footnote{
also at Lab. Fundam. Study, Tomsk State
Pedagogical University, Tomsk}}
\affiliation{
$^1$Department of Physics, Kinki University, Higashi-Osaka 577-8502, Japan\\
$^2$Instituci\`{o} Catalana de Recerca i Estudis Avan\c{c}ats (ICREA)
and Institut de Ciencies de l'Espai (IEEC-CSIC),
Campus UAB, Facultat de Ciencies, Torre C5-Par-2a pl, E-08193 Bellaterra
(Barcelona), Spain
}


\begin{abstract}
We study inflation and late-time acceleration in the expansion
of the universe in non-minimal electromagnetism, in which
the electromagnetic field couples to the scalar curvature function.
It is shown that power-law inflation can be realized due to the non-minimal
gravitational coupling of the electromagnetic field,
and that large-scale magnetic fields can be generated due to the breaking of
the conformal invariance of the electromagnetic field through its non-minimal
gravitational coupling.
Furthermore, it is demonstrated that both inflation and the late-time
acceleration of the universe can be realized in a modified Maxwell-$F(R)$
gravity which is consistent with solar system tests and cosmological bounds
and free of instabilities. At small curvature typical for current universe
the standard Maxwell theory is recovered.
We also consider classically equivalent form of non-minimal
Maxwell-$F(R)$ gravity, and
propose the origin of the non-minimal
gravitational coupling function based on renormalization-group considerations.
\end{abstract}

\preprint{KU-TP\ 019}

\maketitle

\section{Introduction}

It is observationally confirmed not only that inflation occurred
in the early universe, but that the current expansion of the universe is
accelerating~\cite{WMAP1, SN1}.
Although there exist various scenarios to account for the late-time
acceleration in the expansion of the universe, the mechanism
is not well established yet
(for recent reviews, see~\cite{Peebles:2002gy, Padmanabhan:2002ji,
Copeland:2006wr, Durrer:2007re, NO-rev}).

The scenarios to explain the late-time acceleration of the universe
fall into two broad categories~\cite{Durrer:2007re}.
One is general relativistic approaches, i.e., dark energy.
The other is modified gravity approaches, i.e., dark gravity.
As the most promising one of the latter approaches,
the modifications to the Einstein-Hilbert action, e.g.,
the addition of an arbitrary function of the scalar curvature to it,
have been studied (for a review, see~\cite{NO-rev}).
Such a modified theory is considered as an alternative gravitational theory,
so that it must pass cosmological bounds and solar system
tests.

Recently, Hu and Sawicki have proposed a very realistic modified gravitational
theory that evade solar-system tests~\cite{Hu:2007nk}
(for related studies, see~\cite{related studies}).
In this theory, an effective epoch described by the cold dark matter model with cosmological constant ($\Lambda$CDM), which explains high-precision 
observational data, is realized as in general relativity with
cosmological constant (for a review of observational data confronted with
modified gravity, see~\cite{Capozziello:2007ec}).
Although this theory is successful in explaining the late-time acceleration
of the universe, the possibility of the realization of inflation has not
been discussed in Ref.~\cite{Hu:2007nk}.
In Refs.~\cite{Nojiri:2007as,NO1,Eli}, therefore,
modified gravities in which both inflation and the late-time acceleration
of the universe can be realized,
following the previous inflation-acceleration
unification proposal~\cite{Nojiri:2003ft}, have been presented and
investigated. The classification of viable $F(R)$ gravities maybe
suggested too \cite{Eli}.
Here, $F(R)$ is an arbitrary function of the scalar curvature $R$.

As another gravitational source of inflation and the late-time acceleration of
the universe, a coupling between the scalar curvature
and matter Lagrangian has been studied~\cite{matter-1, Allemandi:2005qs}.
Such a coupling may be applied for the realization of the dynamical
cancellation of cosmological constant~\cite{DC}. The criteria for the
viability of such theories have been considered
in Refs.~\cite{criteria-1, Faraoni:2007sn, Bertolami:2007vu}.
Recently, as a simple case, a coupling between the scalar curvature
function
and the kinetic term of a massless scalar field in viable modified
gravity has been
considered~\cite{Nojiri:2007bt}.

On the other hand, it is known that the coupling between the scalar curvature
and the Lagrangian of the electromagnetic field arises in curved spacetime
due to one-loop vacuum-polarization effects in Quantum Electrodynamics
(QED)~\cite{Drummond:1979pp}.
Such a non-minimal gravitational coupling of the electromagnetic field breaks
the conformal invariance of the electromagnetic field, so that
electromagnetic quantum fluctuations can be generated at the inflationary 
stage even in the Friedmann-Robertson-Walker (FRW) spacetime, which
is conformally flat~\cite{Turner:1987bw, Mazzitelli:1995mp, Lambiase:2004zb}.
They can appear as large-scale magnetic fields at the
present time because their scale is made longer due to
inflation.
These large-scale magnetic fields can be the
origin of the large-scale magnetic fields with the field strength
$10^{-7}$--$10^{-6}$G on 10kpc--1Mpc scale observed in clusters of
galaxies~\cite{clusters of galaxies}
(for reviews of cosmic magnetic fields, see~ \cite{magnetic-field-review}).

In the present paper, we consider inflation and the late-time acceleration
of the universe in non-minimal electromagnetism, in which
the electromagnetic field couples to the function of  scalar curvature.
We show that power-law inflation can be realized due to
the non-minimal gravitational coupling of the electromagnetic field,
and that large-scale magnetic fields can be generated due to the breaking of
the conformal invariance of the electromagnetic field through its non-minimal
gravitational coupling\footnote{
In Ref.~\cite{G-M inflation from 5D},
gravitational-electromagnetic inflation from a 5-dimensional vacuum state has
been considered.}. 
The mechanism of inflation in this model is as follows. 
In the very early universe before inflation, electromagnetic quantum 
fluctuations are generated due to the breaking of the conformal invariance of 
the electromagnetic field and they act as a source for inflation. 
Furthermore, also during inflation electromagnetic quantum 
fluctuations are newly generated and the scale is stretched due to inflation, 
so that the scale can be larger than the Hubble horizon at that time, 
and they lead to the large-scale magnetic fields observed in galaxies and 
clusters of galaxies. 
This idea is based on the assumption that 
a given mode is excited quantum mechanically while it is subhorizon sized and 
then as it crosses outside the horizon ``freezes in'' as a classical 
fluctuation~\cite{Turner:1987bw}.
Furthermore, we demonstrate that both inflation and the late-time acceleration
of the universe can be realized in a modified Maxwell-$F(R)$ gravity
proposed in Ref.~\cite{Nojiri:2007as}
which is consistent with solar-system tests and cosmological bounds
and free of instabilities.
We also consider classically equivalent form of non-minimal
Maxwell-$F(R)$ gravity, and
propose the origin of the non-minimal
gravitational coupling function based on renormalization-group considerations.

This paper is organized as follows.
In Sec.\ II we consider a non-minimal gravitational coupling of
the electromagnetic field in general relativity.
First, we describe our model and derive equations of motion from it.
Next, we consider the evolution of the large-scale electric and magnetic
fields. Furthermore, we analyze the gravitational field equation, and then
show that power-law inflation can be realized.
In Sec.\ III we consider a non-minimal gravitational coupling of the
electromagnetic field in a modified gravitational theory proposed
in Ref.~\cite{Nojiri:2007as}. We show that in this theory both inflation and
the late-time acceleration of the universe can be realized.
In Sec.\ IV we consider classically equivalent form of non-minimal
Maxwell-$F(R)$ gravity.
Finally, some summaries are given in Sec.\ V.
In Appendix, we propose the origin of the non-minimal
gravitational coupling function based on renormalization-group considerations.

We use units in which $k_\mathrm{B} = c = \hbar = 1$ and denote the
gravitational constant $8 \pi G$ by ${\kappa}^2$, so that
${\kappa}^2 \equiv 8\pi/{M_{\mathrm{Pl}}}^2$, where
$M_{\mathrm{Pl}} = G^{-1/2} = 1.2 \times 10^{19}$GeV is the Planck mass.
Moreover, in terms of electromagnetism we adopt Heaviside-Lorentz units.

\section{Inflation in general relativity}

In this section, we consider a non-minimal gravitational coupling of
the electromagnetic field in general relativity.

\subsection{Model}

We consider the following model action:
\begin{eqnarray}
S_{\mathrm{GR}} \Eqn{=}
\int d^{4}x \sqrt{-g}
\left[ \hspace{1mm}
{\mathcal{L}}_{\mathrm{EH}}
+{\mathcal{L}}_{\mathrm{EM}}
\hspace{1mm} \right]\,,
\label{eq:2.1} \\[2mm]
{\mathcal{L}}_{\mathrm{EH}}
\Eqn{=} \frac{1}{2\kappa^2} R\,,
\label{eq:2.2} 
\end{eqnarray}
\begin{eqnarray}
{\mathcal{L}}_{\mathrm{EM}}
\Eqn{=}
-\frac{1}{4} I(R)
F_{\mu\nu}F^{\mu\nu}\,,
\label{eq:2.3} \\[2mm]
I(R) 
\Eqn{=} 
1+f(R)\,,
\label{eq:2.4}
\end{eqnarray}
where $g$ is the determinant of the metric tensor $g_{\mu\nu}$,
$R$ is the scalar curvature arising from the spacetime
metric tensor $g_{\mu\nu}$,
and ${\mathcal{L}}_{\mathrm{EH}}$ is the Einstein-Hilbert action.
Moreover, $F_{\mu\nu} = {\partial}_{\mu}A_{\nu} - {\partial}_{\nu}A_{\mu}$
is the electromagnetic field-strength tensor. Here, $A_{\mu}$ is the $U(1)$
gauge field. Furthermore, $f(R)$ is an arbitrary function of $R$.

The field equations can be derived by taking variations of the
action Eq.\ (\ref{eq:2.1}) with respect to the
metric $g_{\mu\nu}$ and the $U(1)$ gauge field $A_{\mu}$ as follows:
\begin{eqnarray}
R_{\mu \nu} - \frac{1}{2}g_{\mu \nu}R
= \kappa^2 T^{(\mathrm{EM})}_{\mu \nu}\,,
\label{eq:2.5}
\end{eqnarray}
with
\begin{eqnarray}
T^{(\mathrm{EM})}_{\mu \nu} 
= 
I(R) \left( g^{\alpha\beta} F_{\mu\beta} F_{\nu\alpha}
-\frac{1}{4} g_{\mu\nu} F_{\alpha\beta}F^{\alpha\beta} \right)
\nonumber 
\end{eqnarray}
\begin{eqnarray}
{}+\frac{1}{2} \biggl\{ f^{\prime}(R)
F_{\alpha\beta}F^{\alpha\beta} R_{\mu \nu}
+ g_{\mu \nu} \Box \left[ f^{\prime}(R)
F_{\alpha\beta}F^{\alpha\beta} \right]
- {\nabla}_{\mu} {\nabla}_{\nu}
\left[ f^{\prime}(R)
F_{\alpha\beta}F^{\alpha\beta} \right]
\biggr\}
\,,
\label{eq:2.6}
\end{eqnarray}
and
\begin{eqnarray}
-\frac{1}{\sqrt{-g}}{\partial}_{\mu}
\left( \sqrt{-g} I(R) F^{\mu\nu}
\right) = 0\,,
\label{eq:2.7}
\end{eqnarray}
where the prime denotes differentiation with respect to $R$,
${\nabla}_{\mu}$ is the covariant derivative operator associated with
$g_{\mu \nu}$, and $\Box \equiv g^{\mu \nu} {\nabla}_{\mu} {\nabla}_{\nu}$
is the covariant d'Alembertian for a scalar field.
In addition, $R_{\mu \nu}$ is the Ricci curvature tensor,
while $T^{(\mathrm{EM})}_{\mu \nu}$ is the contribution to
the energy-momentum tensor from the electromagnetic field.

We assume the spatially flat
Friedmann-Robertson-Walker (FRW) spacetime with the metric
\begin{eqnarray}
{ds}^2 =-{dt}^2 + a^2(t)d{\Vec{x}}^2
= a^2(\eta) ( -{d \eta}^2 + d{\Vec{x}}^2 )\,,
\label{eq:2.8}
\end{eqnarray}
where $a$ is the scale factor, and $\eta$ is the conformal time.
In this spacetime,
$g_{\mu \nu} = \mathrm{diag} \left(-1, a^2(t), a^2(t), a^2(t) \right)$, and
the components of $R_{\mu \nu}$ and $R$ are given by
\begin{eqnarray}
R_{00} = -3\left( \dot{H} + H^2 \right)\,,
\hspace{1mm}
R_{0i} = 0\,,
\hspace{1mm}
R_{ij} = \left( \dot{H} + 3H^2 \right) g_{ij}\,,
\hspace{1mm}
R=6\left( \dot{H} + 2H^2 \right)\,,
\label{eq:2.9}
\end{eqnarray}
where $H=\dot a/a$ is the Hubble parameter. Here, a dot denotes a time
derivative, $\dot{~}=\partial/\partial t$.

\subsection{Evolution of large-scale electric and magnetic fields}

First, we consider the evolution of the $U(1)$ gauge field in this background.
Its equation of motion in the Coulomb gauge ${\partial}^jA_j(t,\Vec{x}) =0$
and the case of $A_0(t,\Vec{x}) = 0$, reads
\begin{eqnarray}
\ddot{A_i}(t,\Vec{x})
+ \left( H + \frac{\dot{I}}{I}
\right) \dot{A_i}(t,\Vec{x})
- \frac{1}{a^2}\Lap\, A_i(t,\Vec{x}) = 0\,,
\label{eq:2.10}
\end{eqnarray}
where $\Lap =  {\partial}^i {\partial}_i$ is the flat 3-dimensional Laplacian.

We shall quantize the $U(1)$ gauge field $A_{\mu}(t,\Vec{x})$.
It follows from the Lagrangian of the electromagnetic field (\ref{eq:2.3})
that the canonical momenta conjugate to $A_{\mu}(t,\Vec{x})$ are given by
\begin{eqnarray}
{\pi}_0 = 0\,, \hspace{5mm} {\pi}_{i} = I a(t) \dot{A_i}(t,\Vec{x})\,.
\label{eq:2.11}
\end{eqnarray}
We impose the canonical commutation relation
between $A_i(t,\Vec{x})$ and ${\pi}_{j}(t,\Vec{x})$,
\begin{eqnarray}
  \left[ \hspace{0.5mm} A_i(t,\Vec{x}), {\pi}_{j}(t,\Vec{y})
  \hspace{0.5mm} \right] = i
 \int \frac{d^3 k}{{(2\pi)}^{3}}
             e^{i \Vecs{k} \cdot \left( \Vecs{x} - \Vecs{y} \right)}
        \left( {\delta}_{ij} - \frac{k_i k_j}{k^2 } \right)\,,
\label{eq:2.12}
\end{eqnarray}
where $\Vec{k}$ is comoving wave number and $k=|\Vec{k}|$. From
this relation, we obtain the expression for $A_i(t,\Vec{x})$ as
\begin{eqnarray}
\hspace{-7mm} A_i(t,\Vec{x}) = \int \frac{d^3 k}{{(2\pi)}^{3/2}}
  \sum_{\sigma=1,2}\left[ \hspace{0.5mm} \hat{b}(\Vec{k},\sigma)
        \epsilon_i(\Vec{k},\sigma)A(k,t)e^{i \Vecs{k} \cdot \Vecs{x} }
       + {\hat{b}}^{\dagger}(\Vec{k},\sigma)
       \epsilon_i^*(\Vec{k},\sigma)
         {A^*}(k,t)e^{-i \Vecs{k} \cdot \Vecs{x}} \hspace{0.5mm} \right],
\label{eq:2.13}
\end{eqnarray}
where $\epsilon_i(\Vec{k},\sigma)$ ($\sigma=1,2$)
are the two orthonormal transverse polarization vectors,
and $\hat{b}(\Vec{k},\sigma)$ and
${\hat{b}}^{\dagger}(\Vec{k},\sigma)$ are the annihilation and creation
operators which satisfy
\begin{eqnarray}
\hspace{-7mm} \left[ \hat{b} ( \Vec{k}, \sigma ),
{\hat{b}}^{\dagger} \left( \tilde{\Vec{k}}, \tilde{\sigma} \right)
 \right] =
\delta_{\sigma, \tilde{\sigma}}
{\delta}^3 \left( \Vec{k}- \tilde{\Vec{k}} \right), \hspace{1mm}
\left[ \hat{b} ( \Vec{k}, \sigma ),
\hat{b} \left( \tilde{\Vec{k}}, \tilde{\sigma} \right)
\right] =
\left[
{\hat{b}}^{\dagger} ( \Vec{k}, \sigma ),
{\hat{b}}^{\dagger} \left( \tilde{\Vec{k}}, \tilde{\sigma} \right)
\right] = 0.
\label{eq:2.14}
\end{eqnarray}
It follows from Eq.~(\ref{eq:2.10}) that the mode function $A(k,t)$
satisfies the equation
\begin{eqnarray}
\ddot{A}(k,t) + \left( H + \frac{\dot{I}}{I} \right)
               \dot{A}(k,t) + \frac{k^2}{a^2} A(k,t) = 0\,,
\label{eq:2.15}
\end{eqnarray}
and that the normalization condition for $A(k,t)$ reads
\begin{eqnarray}
A(k,t){\dot{A}}^{*}(k,t) - {\dot{A}}(k,t){A^{*}}(k,t)
= \frac{i}{I a}\,.
\label{eq:2.16}
\end{eqnarray}
Replacing the independent variable $t$ by $\eta$, we find that
Eq.~(\ref{eq:2.15}) becomes
\begin{eqnarray}
\frac{\partial^2 A(k,\eta)}{\partial \eta^2} +
\frac{1}{I(\eta)} \frac{d I(\eta)}{d \eta}
\frac{\partial A(k,\eta)}{\partial \eta}
+ k^2 {A}(k,\eta) = 0\,.
\label{eq:2.17}
\end{eqnarray}

We are not able to obtain the exact solution of Eq.~(\ref{eq:2.17})
for the case in which $I$ is given by a general function of $\eta$.
In fact, however, we can obtain an approximate solution with sufficient
accuracy by using the Wentzel-Kramers-Brillouin (WKB) approximation on
subhorizon scales and the long-wavelength approximation on superhorizon
scales, and matching these solutions at the horizon
crossing~\cite{Bamba:2006ga, Bamba:2007sx}.

In the exact de Sitter background, we find $-k \eta = k/(aH)$.
Moreover, at the horizon crossing, $H=k/a$ is satisfied, and
hence $-k \eta_k =1$ is satisfied. Here, $\eta_k$ is the conformal time
at the horizon-crossing.
The subhorizon (superhorizon) scale corresponds
to the region $k|\eta|\gg1$ ($k|\eta|\ll1$).
This is expected to be also a sufficiently good definition for
the horizon crossing for power-law inflation $a \propto t^p$, where
$p \gg 1$, which is almost equivalent to exponential inflation because
in this case, $-k \eta = \left[p/(p-1)\right] k/(aH) \approx k/(aH)$.

The WKB subhorizon solution is given by
\begin{eqnarray}
A_{\mathrm{in}} (k,\eta) =
\frac{1}{\sqrt{2k}} I^{-1/2}(\eta) e^{-ik\eta}\,,
\label{eq:2.18}
\end{eqnarray}
where we have assumed that the vacuum in the short-wavelength
limit is the standard Minkowski vacuum.

On the other hand, the solution on superhorizon scales,
$A_{\mathrm{out}} (k,\eta)$, can be obtained
by using the long-wavelength expansion in terms of $k^2$ and
matching this solution with the WKB subhorizon solution in
Eq.~(\ref{eq:2.18}) at the horizon crossing.
The lowest order approximate solution of $A_{\mathrm{out}} (k,\eta)$ is
given by~\cite{Bamba:2006ga}
\begin{eqnarray}
A_{\mathrm{out}} (k,\eta) =
C(k) + D(k) \int_{\eta}^{{\eta}_{\mathrm{f}}}
\frac{1}{I \left( \Tilde{\eta} \right)}
d \Tilde{\eta}\,,
\label{eq:2.19}
\end{eqnarray}
where
\begin{eqnarray}
C(k) \Eqn{=}
\left.
\frac{1}{\sqrt{2k}} I^{-1/2}(\eta)
\left[
1- \left( \frac{1}{2} \frac{d I(\eta)}{d \eta} + i k I(\eta) \right)
\int_{\eta}^{{\eta}_{\mathrm{f}}}
\frac{1}{I \left(\Tilde{\Tilde{\eta}}\right)}
d \Tilde{\Tilde{\eta}} \right] e^{-ik\eta}
\right|_{\eta = \eta_k}\,,
\label{eq:2.20} \\[3mm]
D(k) \Eqn{=}
\left.
\frac{1}{\sqrt{2k}} I^{-1/2}(\eta)
\left( \frac{1}{2} \frac{d I(\eta)}{d \eta} + i k I(\eta) \right)
e^{-ik\eta}
\right|_{\eta = \eta_k}\,.
\label{eq:2.21}
\end{eqnarray}
Neglecting the decaying mode solution,
from Eqs.~(\ref{eq:2.19}) and (\ref{eq:2.20}) we find that
$|A(k,\eta)|^2$ at late times is given by
\begin{eqnarray}
\hspace{-5mm}
\left|A(k,\eta)\right|^2
= |C(k)|^2
= \frac{1}{2kI(\eta_k)}
\left|1- \left[ \frac{1}{2}\frac{1}{kI(\eta_k)}\frac{d I(\eta_k)}{d \eta}
+ i \right]
e^{-ik\eta_k}k\int_{\eta_k}^{{\eta}_{\mathrm{f}}}
\frac{I(\eta_k)}{I \left(\Tilde{\Tilde{\eta}} \right)}
d\Tilde{\Tilde{\eta}}\,\right|^2\,,
\label{eq:2.22}
\end{eqnarray}
where ${\eta}_{\mathrm{f}}$ is the conformal time at the end of inflation.

The proper electric and magnetic fields are given by
\begin{eqnarray}
{E_i}^{\mathrm{proper}}(t,\Vec{x})
\Eqn{=} a^{-1}E_i(t,\Vec{x}) = -a^{-1}\dot{A_i}(t,\Vec{x})\,,
\label{eq:2.23} \\[2mm]
{B_i}^{\mathrm{proper}}(t,\Vec{x})
\Eqn{=} a^{-1}B_i(t,\Vec{x}) = a^{-2}{\epsilon}_{ijk}{\partial}_j
A_k(t,\Vec{x})\,,
\label{eq:2.24}
\end{eqnarray}
where $E_i(t,\Vec{x})$ and $B_i(t,\Vec{x})$ are the comoving electric and
magnetic fields, and ${\epsilon}_{ijk}$ is the totally antisymmetric tensor
(\hspace{0.5mm}${\epsilon}_{123}=1$\hspace{0.5mm}).

Using Eqs.~(\ref{eq:2.19}) and (\ref{eq:2.23}), we find
\begin{eqnarray}
|{E}^{\mathrm{proper}}(k,\eta)|^2
=2 \frac{1}{a^4} \left| \frac{\partial A(k,\eta)}{\partial \eta} \right|^2
=2 \frac{1}{a^4} \frac{|D(k)|^2}{|I(\eta)|^2}\,,
\label{eq:2.25}
\end{eqnarray}
where the factor 2 comes from the two polarization degrees of freedom.
Multiplying $|{E}^{\mathrm{proper}}(k,\eta)|^2$ in Eq.~(\ref{eq:2.25}) by the
phase-space density, $4 \pi k^3/(2\pi)^3$,
we obtain the amplitude of the proper electric fields
in the position space
\begin{eqnarray}
|{E}^{\mathrm{proper}}(L,\eta)|^2 =
\frac{4\pi k^3}{(2\pi)^3}|{E}^{\mathrm{proper}}(k,\eta)|^2
=\frac{|D(k)|^2}{\pi^2 k} \frac{k^4}{a^4} \frac{1}{|I(\eta)|^2}\,,
\label{eq:2.26}
\end{eqnarray}
on a comoving scale $L=2\pi/k$.
Furthermore,
the energy density of the large-scale electric fields
in the position space is given by
\begin{eqnarray}
\rho_E (L,\eta) =
\frac{1}{2} |{E}^{\mathrm{proper}}(L,\eta)|^2 I(\eta)
=\frac{|D(k)|^2}{2 \pi^2 k} \frac{k^4}{a^4} \frac{1}{I(\eta)}\,.
\label{eq:2.27}
\end{eqnarray}

Similarly, using Eqs.~(\ref{eq:2.19}) and (\ref{eq:2.24}), we find
\begin{eqnarray}
|{B}^{\mathrm{proper}}(k,\eta)|^2
=2\frac{k^2}{a^4}|A(k,\eta)|^2
=2\frac{k^2}{a^4}|C(k)|^2\,,
\label{eq:2.28}
\end{eqnarray}
where the factor 2 comes from the two polarization degrees of freedom.
Multiplying $|{B}^{\mathrm{proper}}(k,\eta)|^2$ in Eq.~(\ref{eq:2.28}) by the
phase-space density, $4 \pi k^3/(2\pi)^3$,
we obtain the amplitude of the proper magnetic fields
in the position space
\begin{eqnarray}
|{B}^{\mathrm{proper}}(L,\eta)|^2 =
\frac{4\pi k^3}{(2\pi)^3}|{B}^{\mathrm{proper}}(k,\eta)|^2
= \frac{k|C(k)|^2}{\pi^2}\frac{k^4}{a^4}\,,
\label{eq:2.29}
\end{eqnarray}
on a comoving scale $L=2\pi/k$.
Furthermore,
the energy density of the large-scale magnetic fields
in the position space is given by
\begin{eqnarray}
\rho_B(L,\eta) =
\frac{1}{2} |{B}^{\mathrm{proper}}(L,\eta)|^2 I(\eta)
= \frac{k|C(k)|^2}{2\pi^2}\frac{k^4}{a^4} I(\eta)\,.
\label{eq:2.30}
\end{eqnarray}

\if
Here we note the following point. 
As is stated in the next subsection, we consider the case in which 
the value of $I$ during inflation is much larger than unity and 
the variation of $I$ in time during inflation is much smaller than unity 
for de Sitter (exponential) inflation, namely, the case of 
Eqs.~(\ref{eq:2.41}), (\ref{eq:2.44}) and (\ref{eq:2.45}) with 
$c_1/c_2 \gg 1$. In this case, from Eq.~(\ref{eq:2.20}) we find 
$k|C(k)|^2 \approx 1/\left[ 2I(\eta_k) \right]$. Hence, using this approximate 
relation and Eq.~(\ref{eq:2.30}), we find 
\begin{eqnarray}
\rho_B(L,\eta) \approx 
\frac{1}{\left( 2\pi \right)^2} \left(\frac{k}{a}\right)^4 
\frac{I(\eta)}{I(\eta_k)}\,.
\label{eq:2.31}
\end{eqnarray}
In fact, because the horizon crossing occurs during inflation, it follows from 
Eq.~(\ref{eq:2.31}) that the energy density of the large-scale magnetic fields 
around the horizon crossing during inflation is estimated as 
\begin{eqnarray}
\rho_B(L,\eta) \approx \left( \frac{H^2}{2\pi} \right)^2 \,, 
\label{eq:2.32}
\end{eqnarray}
where we have used $k/a \approx H$ and $I(\eta) \approx I(\eta_k)$. 
\fi

Here we note the following point. 
As an example, if $I$ is given by the following form: 
$
I(\eta)=I_{\mathrm{s}} 
\left( \eta/\eta_{\mathrm{s}} \right)^{-\alpha},
$
where $\eta_{\mathrm{s}}$ is some fiducial time during inflation, 
$I_{\mathrm{s}}$ is the value of $I(\eta)$ at $\eta=\eta_{\mathrm{s}}$, 
and $\alpha$ is a constant, from Eq.~(\ref{eq:2.20}) we find 
$k|C(k)|^2 = {\cal C}/\left[ 2I(\eta_k) \right]$, where 
${\cal C}$ is a constant of order unity~\cite{Bamba:2006ga, Bamba:2007sx}. 
Hence, using this relation and Eq.~(\ref{eq:2.30}), we find 
\begin{eqnarray}
\rho_B(L,\eta) = 
\frac{{\cal C}}{\left( 2\pi \right)^2} \left(\frac{k}{a}\right)^4 
\frac{I(\eta)}{I(\eta_k)}\,.
\label{eq:2.31}
\end{eqnarray}

\subsection{Power-law inflation}

The $(\mu,\nu)=(0,0)$ component and
the trace part of the $(\mu,\nu)=(i,j)$ component of Eq.~(\ref{eq:2.5}),
where $i$ and $j$ run from $1$ to $3$, read
\begin{eqnarray}
\hspace{-10mm}
H^2 + J_1 
\Eqn{=}
\frac{\kappa^2}{3}
\biggl\{
I(R) \left( g^{\alpha\beta} F_{0\beta} F_{0\alpha}
-\frac{1}{4} g_{00} F_{\alpha\beta}F^{\alpha\beta} \right)
\nonumber \\[2mm]
&&\hspace{0mm}
{}+ \frac{3}{2} \left[
-f^{\prime}(R) \left( \dot{H} + H^2 \right) +
6 f^{\prime\prime}(R) H \left( \ddot{H} + 4H\dot{H} \right)
\right] F_{\alpha\beta}F^{\alpha\beta}
\nonumber \\[2mm]
&&\hspace{0mm}
{}+ \frac{3}{2} f^{\prime}(R) H \left( F_{\alpha\beta}F^{\alpha\beta}
\right)^{\bullet}
-\frac{1}{2} f^{\prime}(R) \frac{1}{a^2}
\Lap \left( F_{\alpha\beta}F^{\alpha\beta} \right)
\biggr\}\,,
\label{eq:2.32} \\[2mm]
J_1 \Eqn{=} 
\frac{1}{6} F(R)  - F^{\prime}(R) \left( \dot{H} + H^2 \right)\,,
\label{eq:2.33}
\end{eqnarray}
and
\begin{eqnarray}
&&
2\dot{H} + 3H^2 + J_2 
\nonumber \\[2mm]
&& \hspace{5mm}
=\frac{\kappa^2}{2}
\biggl\{
\frac{1}{6} I(R) F_{\alpha\beta}F^{\alpha\beta}
+ \biggl[
-f^{\prime}(R) \left( \dot{H} + 3H^2 \right)
\nonumber \\[2mm]
&& \hspace{5mm}
{}+6f^{\prime\prime}(R)
\left(\dddot{H}+7H\ddot{H}+4\dot{H}^2+12H^2\dot{H} \right)
+36f^{\prime\prime\prime}(R) \left( \ddot{H}+4H\dot{H} \right)^2
\biggr] F_{\alpha\beta}F^{\alpha\beta}
\nonumber \\[2mm]
&& \hspace{5mm}
{}+3\left[f^{\prime}(R)H + 4f^{\prime\prime}(R)
\left( \ddot{H}+4H\dot{H} \right) \right]
\left( F_{\alpha\beta}F^{\alpha\beta} \right)^{\bullet}
+ f^{\prime}(R) \left( F_{\alpha\beta}F^{\alpha\beta} \right)^{\bullet\bullet}
\nonumber \\[2mm]
&& \hspace{5mm}
{}-\frac{2}{3}f^{\prime}(R) \frac{1}{a^2}
\Lap \left( F_{\alpha\beta}F^{\alpha\beta} \right)
\biggr\}\,,
\label{eq:2.34} \\[2mm]
&& \hspace{1mm}
J_2 =
\frac{1}{2}F(R)
- F^{\prime}(R) \left( \dot{H} + 3H^2 \right)
+ 6F^{\prime\prime}(R) \left[
\dddot{H} + 4\left( \dot{H}^2 + H\ddot{H} \right)
\right]
\nonumber \\[2mm]
&& \hspace{10mm}
{}+ 36F^{\prime\prime\prime}(R) \left( \ddot{H} + 4H\dot{H} \right)^2\,,
\label{eq:2.35}
\end{eqnarray}
where
\begin{eqnarray}
\hspace{-15mm}
g^{\alpha\beta} F_{0\beta} F_{0\alpha}
-\frac{1}{4} g_{00} F_{\alpha\beta}F^{\alpha\beta}
\Eqn{=} \frac{1}{2}
\left( |{E_i}^{\mathrm{proper}}(t,\Vec{x})|^2
+ |{B_i}^{\mathrm{proper}}(t,\Vec{x})|^2 \right)\,,
\label{eq:2.36} \\[2mm]
F_{\alpha\beta}F^{\alpha\beta}
\Eqn{=} 2 \left(|{B_i}^{\mathrm{proper}}(t,\Vec{x})|^2
- |{E_i}^{\mathrm{proper}}(t,\Vec{x})|^2 \right)\,,
\label{eq:2.37}
\end{eqnarray}
respectively.
Here, a large dot in terms of $F_{\alpha\beta}F^{\alpha\beta}$
denotes a time derivative,
$\left( F_{\alpha\beta}F^{\alpha\beta} \right)^{\bullet}
= \partial \left( F_{\alpha\beta}F^{\alpha\beta} \right) /\partial t$. 
Moreover, $J_1$ and $J_2$ are correction terms in 
a modified gravitational theory described by the action 
in Eq.~(\ref{eq:3.1}) in 
the next section. Hence, because in this section we consider 
general relativity, i.e., the case $F(R)=0$ in the action in 
Eq.~(\ref{eq:3.2}), 
here both $J_1$ and $J_2$ are zero. 
In deriving Eqs.~(\ref{eq:2.32}) and (\ref{eq:2.34}), we have used
equations in (\ref{eq:2.9}). Moreover, 
in deriving Eqs.~(\ref{eq:2.36}) and (\ref{eq:2.37}), we have used
Eqs.~(\ref{eq:2.23}) and (\ref{eq:2.24}). Furthermore,
applying Eqs.~(\ref{eq:2.26}) and (\ref{eq:2.29}) to
$|{E_i}^{\mathrm{proper}}(t,\Vec{x})|^2$ and
$|{B_i}^{\mathrm{proper}}(t,\Vec{x})|^2$, respectively, we find
\begin{eqnarray}
\left( F_{\alpha\beta}F^{\alpha\beta} \right)^{\bullet}
\Eqn{=} 8\biggl\{
- H |{B}^{\mathrm{proper}}(L,\eta)|^2
\nonumber \\[2mm]
&&
{}+\left[
H + 3\frac{f^{\prime}(R) }{1+f(R)} \left( \ddot{H} + 4H\dot{H} \right)
\right] |{E}^{\mathrm{proper}}(L,\eta)|^2
\biggr\}\,.
\label{eq:2.38}
\end{eqnarray}

Here we consider the case in which magnetic fields are mainly generated
rather than electric fields because we are interested in the generation of
large-scale magnetic fields.
It follows from Eqs.~(\ref{eq:2.27}) and (\ref{eq:2.30}) that this situation 
is realized if $I$ increases rapidly in time during 
inflation~\cite{Bamba:2007sx}. 
(Hence, from this point we neglect terms in electric fields.)
Moreover, we consider the case in which 
$\Lap \left( F_{\alpha\beta}F^{\alpha\beta} \right)$ is very small because it
corresponds to the second order spatial derivative of the quadratic quantity of electromagnetic quantum fluctuations, so that it can be neglected.
In this case, using Eqs.~(\ref{eq:2.29}) and (\ref{eq:2.38}), we find that
Eqs.~(\ref{eq:2.32}) and (\ref{eq:2.34}) are reduced to
\begin{eqnarray}
H^2
= \kappa^2
\left[
\frac{1}{6} I(R)
-f^{\prime}(R) \left( \dot{H} + 5H^2 \right) +
6 f^{\prime\prime}(R) H \left( \ddot{H} + 4H\dot{H} \right)
\right]
\frac{k|C(k)|^2}{\pi^2}\frac{k^4}{a^4}\,,
\label{eq:2.39}
\end{eqnarray}
and
\begin{eqnarray}
2\dot{H} + 3H^2
\Eqn{=} \kappa^2
\biggl[
\frac{1}{6} I(R)
+f^{\prime}(R) \left( - 5\dot{H} + H^2 \right) +
6 f^{\prime\prime}(R) \left( \dddot{H} -H\ddot{H}+4\dot{H}^2-20H^2\dot{H}
\right)
\nonumber \\[2mm]
&&
{}+36f^{\prime\prime\prime}(R)
\left( \ddot{H} + 4H\dot{H} \right)^2
\biggr]
\frac{k|C(k)|^2}{\pi^2}\frac{k^4}{a^4}\,,
\label{eq:2.40}
\end{eqnarray}
respectively.
Eliminating $I(R)$ from Eqs.~(\ref{eq:2.39}) and (\ref{eq:2.40}),
we obtain
\begin{eqnarray}
\dot{H} + H^2
\Eqn{=} \kappa^2
\biggl[
f^{\prime}(R) \left( -2\dot{H} + 3H^2 \right) +
3f^{\prime\prime}(R) \left( \dddot{H}-2H\ddot{H}+4\dot{H}^2-24H^2\dot{H}
\right)
\nonumber \\[2mm]
&&
{}+18f^{\prime\prime\prime}(R)
\left( \ddot{H} + 4H\dot{H} \right)^2
\biggr]
\frac{k|C(k)|^2}{\pi^2}\frac{k^4}{a^4}\,.
\label{eq:2.41}
\end{eqnarray}

Here we consider the case in which $f(R)$ is given by the following form:
\begin{eqnarray}
f(R) = f_{\mathrm{HS}}(R) \equiv \frac{c_1 \left(R/m^2 \right)^n}
{c_2 \left(R/m^2 \right)^n + 1}\,,
\label{eq:2.42}
\end{eqnarray}
which satisfies the conditions:
\begin{eqnarray}
\lim_{R\to\infty} f_{\mathrm{HS}}(R)
\Eqn{=} \frac{c_1}{c_2} = \mbox{const}\,,
\label{eq:2.43} \\[2mm]
\lim_{R\to 0} f_{\mathrm{HS}}(R)
\Eqn{=} 0\,.
\label{eq:2.44}
\end{eqnarray}
Here, $c_1$ and $c_2$ are dimensionless constants, $n$ is a positive
constant, and $m$ denotes a mass scale. This form, $f_{\mathrm{HS}}(R)$, has
been proposed by Hu and Sawicki~\cite{Hu:2007nk}.
The second condition (\ref{eq:2.44}) means that there could exist a flat
spacetime solution.
Hence, because in the late time universe the value of the scalar curvature
becomes zero, the electromagnetic coupling $I$ becomes unity, so
that the ordinary Maxwell theory can be naturally recovered.

In order to show that power-law inflation can be realized,
we consider the case in which the scale factor is given by 
$a(t) = \bar{a} \left(t/\bar{t}\right)^p$, 
where $\bar{t}$ is some fiducial time during inflation, 
$\bar{a}$ is the value of $a(t)$ at $t=\bar{t}$, 
and $p$ is a positive constant. 
In this case, $H=p/t$, $\dot{H}=-p/t^2$,
$\ddot{H}=2p/t^3$, and $\dddot{H}=-6p/t^4$. Moreover, it follows from
the fourth equation in (\ref{eq:2.9}) that $R=6p(2p-1)/t^2$.
At the inflationary stage, because $R/m^2 \gg 1$, we are able to
use the following approximate relations:
\begin{eqnarray}
f_{\mathrm{HS}}(R) \Eqn{\approx} \frac{c_1}{c_2} \left[1-\frac{1}{c_2}
\left( \frac{R}{m^2} \right)^{-n} \right]\,,
\label{eq:2.45} \\[2mm]
f_{\mathrm{HS}}^{\prime}(R) \Eqn{\approx}
\frac{nc_1}{{c_2}^2} \frac{1}{m^2} \left( \frac{R}{m^2} \right)^{-(n+1)}\,,
\label{eq:2.46} \\[2mm]
f_{\mathrm{HS}}^{\prime\prime}(R) \Eqn{\approx}
-\frac{n(n+1)c_1}{{c_2}^2} \frac{1}{m^4} \left( \frac{R}{m^2}
\right)^{-(n+2)}\,,
\label{eq:2.47} \\[2mm]
f_{\mathrm{HS}}^{\prime\prime\prime}(R) \Eqn{\approx}
\frac{n(n+1)(n+2)c_1}{{c_2}^2} \frac{1}{m^6} \left( \frac{R}{m^2}
\right)^{-(n+3)}\,.
\label{eq:2.48}
\end{eqnarray}
Substituting the above relations in terms of $a$, $H$, and $R$ and
Eqs.~(\ref{eq:2.46})--(\ref{eq:2.48}) into Eq.~(\ref{eq:2.41}), we find
\begin{eqnarray}
p \Eqn{=} \frac{n+1}{2}\,, 
\label{eq:2.49} \\[2mm] 
\frac{\bar{a}}{\bar{t}^p} \Eqn{=} 
\left\{
\frac{1}{3^{n+1}\pi^2} \frac{1}{(n-1) \left[ n(n+1) \right]^n} 
\frac{(-c_1)}{{c_2}^2} k|C(k)|^2 k^4 \kappa^2 m^{2n} \right\}^{1/4}\,.
\label{eq:2.50}
\end{eqnarray}
Hence, if $n \gg 1$, $p$ becomes much larger than unity, so that
power-law inflation can be realized.
Consequently, it follows form this result that
the electromagnetic field with a non-minimal gravitational
coupling in Eq.~(\ref{eq:2.3}) can be a source of inflation.

Here we state the two following points. 
In this paper we consider only the case in which the values of 
the terms proportional to 
$f^{\prime}(R)$, $f^{\prime\prime}(R)$ and $f^{\prime\prime\prime}(R)$ 
in the right-hand side of 
Eqs.~(\ref{eq:2.39}) and (\ref{eq:2.40}) are dominant to the value of 
the term proportional to $I(R)$. 
Among the terms proportional to 
$f^{\prime}(R)$, $f^{\prime\prime}(R)$ and $f^{\prime\prime\prime}(R)$, 
the term proportional to $f^{\prime}(R)$ is dominant, and its value is 
order $f^{\prime}(R) H^2 \approx n \left(c_1/{c_2}^2\right) 
\left( H^2/m^2 \right) \left( R/m^2 \right)^{-(n+1)}$, which follows from 
Eq.~(\ref{eq:2.46}). 
Here, it follows from $H=p/t$ and $R=6p(2p-1)/t^2$ that $R$ is order $10 H^2$. 
The condition that the term proportional to $f^{\prime}(R)$ is 
dominant in the source term would be 
$I(R)/\left[ f^{\prime}(R) H^2 \right] \sim 10 c_2 (R/m^2)^n/n \ll 1$. 
This would require extremely small $c_2$ because at the inflationary stage 
$R/m^2 \gg 1$ and $n \gg 1$. 
In such a case, the value of the right-hand side of Eq.~(\ref{eq:2.41}), 
which is order $\kappa^2 f^{\prime}(R) H^2 \rho_B/I$ 
(this estimation is derived by using Eq.~(\ref{eq:2.30})), 
can be order $H^2$. 
Consequently, the right-hand side of Eq.~(\ref{eq:2.41}) can balance with 
the left-hand side of Eq.~(\ref{eq:2.41}), and hence Eq.~(\ref{eq:2.41}) 
can be satisfied without contradiction to the result, i.e., 
power-law inflation in which $p$ is much larger than unity can be realized. 
The reason why we consider the case in which 
the term proportional to $I(R)$ on the right-hand side of 
Eqs.~(\ref{eq:2.39}) and (\ref{eq:2.40}) is so small in comparison with 
the term proportional to $f^{\prime}(R)$ that it can be neglected is 
as follows: 
If the opposite case, namely, 
the term proportional to $I(R)$ is dominant to the term proportional to 
$f^{\prime}(R)$, Eqs.~(\ref{eq:2.39}) and (\ref{eq:2.40}) are approximately written as $H^2 \approx \kappa^2 \rho_B/6$ and $2\dot{H}+3H^2 
\approx \kappa^2 \rho_B/6$, respectively. 
Thus, in this case it follows from Eqs.~(\ref{eq:2.39}) and (\ref{eq:2.40}) 
that $H^2$ and $2\dot{H}+3H^2$ are the same order and their difference, 
$2\dot{H}+2H^2$, must be much smaller than $H^2$. In fact, 
Eq.~(\ref{eq:2.41}) implies that $\dot{H}+H^2$ balances with much smaller 
quantity than $\kappa^2 \rho_B$. Now, $\left(\dot{H}+H^2 \right)/H^2 = 
(p-1)/p$ and hence $p$ must be very close to unity. Consequently, in this case 
power-law inflation cannot be realized. 

Furthermore, 
when we consider the non-minimal electromagnetic theory described by 
Eq.~(\ref{eq:2.3}), in the very early universe before the beginning of 
inflation electromagnetic quantum fluctuations can be generated due to 
the breaking of the conformal invariance of the electromagnetic field through 
its non-minimal gravitational coupling, This is because it is considered that 
in the very early universe before inflation 
(e.g., the grand unified theory (GUT) scale), 
there can exist quantum fluctuations of all physical 
quantities, as the quantum fluctuations of the inflaton field 
in the chaotic inflation scenario~\cite{Linde:1983gd}.
On the other hand, the non-minimal coupling between the electromagnetic field 
and the scalar curvature function $f(R)$ is purely classical. 
Furthermore, as explained above, in this paper we consider the case 
in which the term proportional to $f^{\prime}(R)$ in the right-hand side 
of Eqs.~(\ref{eq:2.39}) and (\ref{eq:2.40}) is dominant to the term 
proportional to $I(R)$. 
Hence, power-law inflation can be realized due to not 
the term proportional to $I(R)$, namely, the energy density of large-scale 
magnetic fields, but the term proportional to $f^{\prime}(R)$, namely, 
a non-minimal electromagnetic coupling. 
Consequently, in this model we consider that inflation can be realized due to not purely quantum effects but semi-classical effects. 

Finally, we note the following three points.
In the present model, large-scale magnetic fields can be generated
due to the breaking of the conformal
invariance of the electromagnetic field through a coupling with the scalar
curvature, $I(R) F_{\mu\nu}F^{\mu\nu}$, as is shown in the preceding
subsection. If there does not exist such a coupling, i.e., $f(R)=0$ and
hence $I=1$, in which the ordinary Maxwell theory is realized, electromagnetic
quantum fluctuations cannot be generated in the FRW spacetime
because this background spacetime is conformally flat. 
This result is also realized in the case of dilaton
electromagnetism~\cite{Ratra:1991bn, Lemoine:1995vj, Gasperini:1995dh,
Bamba-mag1, Martin:2007ue}, 
in which 
the Lagrangian of the electromagnetic field is given by
$\tilde{I}(\Phi) F_{\mu\nu}F^{\mu\nu}$ with
$\tilde{I}(\Phi)=e^{\lambda\kappa\Phi}$~\cite{Bamba-mag1},
where $\Phi$ is the dilaton field and $\lambda$ is a dimensionless
constant. (It is also realized in other scalar-field
electromagnetism~\cite{Giovannini-mag, Garretson:1992vt, Field:1998hi}.)

Moreover, Bertolami and P\'aramos have recently considered constraints
on a non-minimal gravitational coupling of matter, namely, for the present
model, $f(R)$ in Eq.~(\ref{eq:2.4}), from the observational data of the
central temperature of the Sun~\cite{Bertolami:2007vu}.
They have studied the effect of a non-minimal gravitational coupling of
matter on the hydrostatic equilibrium of the spherically symmetric system with
a polytropic equation of state approximately describing the Sun with
sufficient accuracy,
assuming a perturbative regime to the usual Tolman-Oppenheimer-Volkoff (TOV)
equation of hydrostatic equilibrium and taking into account the validity
of the Newtonian regime in a theory with a non-minimal gravitational coupling
of matter.
According to them, there exists no strong constraints on a
non-minimal gravitational coupling of matter obtained from the comparison of
the predictions of the theoretical models and the current
observational sensitivity to the central temperature of the Sun
except for the relation of the perturbative approach, $|f(R)| \ll 1$.
It follows from the relation $|f(R)| \ll 1$ that for the case
$f(R) = f_{\mathrm{HS}}(R)$
in Eq.~(\ref{eq:2.42}) with $m=m_e=0.511 \mathrm{MeV}$~\cite{Drummond:1979pp},
where $m_e$ is the electron mass,
using the maximum value of the central mass density of the Sun,
$\rho_{\mathrm{c}} = 1.62 \times 10^2\, \mathrm{g}\, \mathrm{cm}^{-3}$, and
the expression of the scalar curvature in the Newtonian regime,
$R \approx -8 \pi G \rho_{\mathrm{c}}$, we find that
the constraint on $f_{\mathrm{HS}}(R)$ is given by
\begin{eqnarray}
\left| f_{\mathrm{HS}}(R) \right| \approx
\left| \frac{c_1 \left( -8 \pi G \rho_{\mathrm{c}}/m_e^2 \right)^n}
{c_2 \left( -8 \pi G \rho_{\mathrm{c}}/m_e^2 \right)^n + 1} \right| =
\left| \frac{c_1 \left( -4.51 \times 10^{-46} \right)^n}
{c_2 \left( -4.51 \times 10^{-46} \right)^n + 1} \right| \ll 1\,.
\label{eq:2.51}
\end{eqnarray}

Furthermore, the existence of the non-minimal gravitational coupling of
the electromagnetic field $f(R)$ in Eq.~(\ref{eq:2.4}) changes the
value of the fine structure constant, i.e., the strength of the
electromagnetic coupling. Hence, the deviation of the non-minimal
electromagnetism from the usual Maxwell theory can be constrained
from the observations of radio and optical quasar absorption
lines~\cite{Tzanavaris:2006uf}, those of the anisotropy of the cosmic
microwave background (CMB) radiation~\cite{Battye:2000ds, Stefanescu:2007aa},
those of the absorption of CMB radiation at 21 cm hyperfine transition of
the neutral atomic hydrogen~\cite{Khatri:2007yv},
and big bang nucleosynthesis (BBN)~\cite{Bergstrom:1999wm, Avelino:2001nr}
as well as solar-system experiments~\cite{Fujii:2006ic}
(for a recent review, see~\cite{GarciaBerro:2007ir}).

\section{Inflation and late-time cosmic acceleration in modified gravity}

In this section, we consider a non-minimal gravitational coupling of the electromagnetic field in a modified gravitational theory proposed
in Ref.~\cite{Nojiri:2007as}.

\subsection{Inflation}

We consider the following model action:
\begin{eqnarray}
S_{\mathrm{MG}}
\Eqn{=}
\int d^{4}x \sqrt{-g}
\left[ \hspace{1mm}
{\mathcal{L}}_{\mathrm{MG}}
+{\mathcal{L}}_{\mathrm{EM}}
\hspace{1mm} \right]\,,
\label{eq:3.1} \\[2mm] 
{\mathcal{L}}_{\mathrm{MG}}
\Eqn{=}
\frac{1}{2\kappa^2} \left[ R+F(R) \right]\,,
\label{eq:3.2}
\end{eqnarray}
where $F(R)$ is an arbitrary function of $R$.
Here, ${\mathcal{L}}_{\mathrm{EM}}$ is given by Eq.~(\ref{eq:2.3}).
We note that $F(R)$ is the modified part of gravity, and hence
$F(R)$ is completely different from the non-minimal gravitational
coupling of the electromagnetic field $f(R)$ in Eq.\ (\ref{eq:2.4}).

Taking variations of the
action Eq.\ (\ref{eq:3.1}) with respect to the
metric $g_{\mu\nu}$, we find that the field equation of modified gravity
is given by~\cite{Nojiri:2007as}
\begin{eqnarray}
\left[ 1+F^{\prime}(R) \right] R_{\mu \nu}
- \frac{1}{2}g_{\mu \nu} \left[ R+F(R) \right] + g_{\mu \nu}
\Box F^{\prime}(R) - {\nabla}_{\mu} {\nabla}_{\nu} F^{\prime}(R)
= \kappa^2 T^{(\mathrm{EM})}_{\mu \nu}\,.
\label{eq:3.3}
\end{eqnarray}

The $(\mu,\nu)=(0,0)$ component and 
the trace part of the $(\mu,\nu)=(i,j)$ component of Eq.~(\ref{eq:3.3}), 
where $i$ and $j$ run from $1$ to $3$, are given by 
Eqs.~(\ref{eq:2.32}) and (\ref{eq:2.34}), respectively. 
Similarly to the preceding section, 
we here consider the case in which 
terms in electric fields and 
$\Lap \left( F_{\alpha\beta}F^{\alpha\beta} \right)$ are negligible.
In this case, 
eliminating $I(R)$ from Eqs.~(\ref{eq:2.32}) and (\ref{eq:2.34}), 
we obtain
\begin{eqnarray}
&& \hspace{-5mm}
\dot{H} + H^2 +
\biggl\{ \frac{1}{6} F(R) - F^{\prime}(R) H^2 +
3F^{\prime\prime}(R) \left[
\dddot{H} + 4\left( \dot{H}^2 + H\ddot{H} \right) \right]
+ 18F^{\prime\prime\prime}(R) \left( \ddot{H} + 4H\dot{H} \right)^2
\biggr\}
\nonumber \\[2mm]
&& \hspace{5mm}
{}= \kappa^2
\biggl[
f^{\prime}(R) \left( -2\dot{H} + 3H^2 \right) +
3f^{\prime\prime}(R) \left( \dddot{H}-2H\ddot{H}+4\dot{H}^2-24H^2\dot{H}
\right)
\nonumber \\[2mm]
&& \hspace{5mm}
{}+18f^{\prime\prime\prime}(R)
\left( \ddot{H} + 4H\dot{H} \right)^2
\biggr]
\frac{k|C(k)|^2}{\pi^2}\frac{k^4}{a^4}\,.
\label{eq:3.4}
\end{eqnarray}

Here we consider the case in which $F(R)$ is given by
\begin{eqnarray}
F(R) \Eqn{=}
- M^2 \frac{\left[ \left(R/M^2\right) - \left(R_0/M^2\right) \right]^{2l+1}
+ {\left(R_0/M^2\right)}^{2l+1}}
{c_3 + c_4 \left\{
\left[ \left(R/M^2\right) - \left(R_0/M^2\right) \right]^{2l+1} +
{\left(R_0/M^2\right)}^{2l+1} \right\}}\,,
\label{eq:3.5}
\end{eqnarray}
which satisfies the following conditions: 
$
\lim_{R\to\infty} F(R) = -M^2/c_4 = \mbox{const}, 
$
$
\lim_{R\to 0} F(R) = 0. 
$
Here, $c_3$ and $c_4$ are dimensionless constants,
$l$ is a positive integer, and $M$ denotes a mass scale.
We consider that in the limit $R\to\infty$, i.e., at the very early stage of 
the universe, $F(R)$ becomes an effective cosmological constant, and that
at the present time $F(R)$ becomes a small constant, namely,
\begin{eqnarray}
\lim_{R\to\infty} F(R)
\Eqn{=} -M^2\frac{1}{c_4} = -2{\Lambda}_{\mathrm{i}}\,,
\label{eq:3.6}
\end{eqnarray}
\begin{eqnarray}
F(R_0)
\Eqn{=} -M^2 \frac{\left(R_0/M^2\right)^{2l+1}}
{c_3 + c_4 \left(R_0/M^2\right)^{2l+1}} = -2R_0\,,
\label{eq:3.7}
\end{eqnarray}
where ${\Lambda}_{\mathrm{i}} \left(\gg {H_0}^2 \right)$ is an effective
cosmological constant in the very early universe and
$R_0 \left(\approx {H_0}^2 \right)$ is current curvature. Here,
$H_0$ is the Hubble constant at the present time~\cite{Kolb and Turner}:
$H_{0} = 100 h \hspace{1mm} \mathrm{km} \hspace{1mm} {\mathrm{s}}^{-1}
\hspace{1mm} {\mathrm{Mpc}}^{-1}
= 2.1 h \times 10^{-42} {\mathrm{GeV}}
\approx 1.5 \times 10^{-33} {\mathrm{eV}}$, where
we have used $h=0.70$~\cite{Freedman:2000cf}. From
Eqs.~(\ref{eq:3.6}) and (\ref{eq:3.7}), we find
\begin{eqnarray}
c_3 \Eqn{=} \frac{1}{2} \left( \frac{R_0}{M^2} \right)^{2l}
\left( 1- \frac{R_0}{{\Lambda}_{\mathrm{i}}} \right)
\approx \frac{1}{2} \left( \frac{R_0}{M^2} \right)^{2l}\,,
\label{eq:3.8} \\[2mm]
c_4 \Eqn{=} \frac{1}{2} \frac{M^2}{{\Lambda}_{\mathrm{i}}}\,,
\label{eq:3.9}
\end{eqnarray}
where the last approximate equality in Eq.~(\ref{eq:3.8}) follows from
$\left( R_0/{\Lambda}_{\mathrm{i}} \right) \ll 1$.

Furthermore, we consider the case in which $f(R)$ is given by the
following form:
\begin{eqnarray}
f(R) = f_{\mathrm{NO}}(R) \equiv
\frac{\left[ \left(R/M^2\right) - \left(R_0/M^2\right) \right]^{2q+1}
+ {\left(R_0/M^2\right)}^{2q+1}}
{c_5 + c_6 \left\{
\left[ \left(R/M^2\right) - \left(R_0/M^2\right) \right]^{2q+1} +
{\left(R_0/M^2\right)}^{2q+1} \right\}}\,,
\label{eq:3.10}
\end{eqnarray}
which satisfies the following conditions: 
$
\lim_{R\to\infty} f_{\mathrm{NO}}(R) = 1/c_6 = \mbox{const}, 
$
$
\lim_{R\to 0} f_{\mathrm{NO}}(R) = 0. 
$
Here, $c_5$ and $c_6$ are dimensionless constants, and 
$q$ is a positive integer.
The form of $F(R)$ in Eq.~(\ref{eq:3.5}) and
$f_{\mathrm{NO}}(R)$ in Eq.~(\ref{eq:3.10})
is taken from Ref.~\cite{Nojiri:2007as}. This 
form corresponds to the extension of the form of $f_{\mathrm{HS}}(R)$ in
Eq.~(\ref{eq:2.42}). It has been shown in Ref.~\cite{Nojiri:2007as}
that modified gravitational theories described by
the action (\ref{eq:3.2}) with $F(R)$ in Eq.~(\ref{eq:3.5}) successfully
pass the solar-system tests as well as cosmological bounds
and they are free of instabilities.

At the inflationary stage, because $R/M^2 \gg 1$ and $R/M^2 \gg R_0/M^2$,
we are able to use the following approximate relations:
\begin{eqnarray}
F(R) \Eqn{\approx}
-M^2 \frac{1}{c_4} \left[1-\frac{c_3}{c_4}
\left( \frac{R}{M^2} \right)^{-(2l+1)} \right]\,,
\label{eq:3.11} 
\end{eqnarray}
and
\begin{eqnarray}
f_{\mathrm{NO}}(R) \Eqn{\approx}
\frac{1}{c_6} \left[1-\frac{c_5}{c_6}
\left( \frac{R}{M^2} \right)^{-(2q+1)} \right]\,.
\label{eq:3.12} 
\end{eqnarray}

At the very early stage of the universe, because $R\to\infty$, it follows from 
Eq.~(\ref{eq:3.6}) and the condition, 
$
\lim_{R\to\infty} f_{\mathrm{NO}}(R) = 1/c_6 = \mbox{const}, 
$
that Eq.~(\ref{eq:3.4}) are reduced to 
\begin{eqnarray}
\dot{H} + H^2 = \frac{{\Lambda}_{\mathrm{i}}}{3}\,.
\label{eq:3.13}
\end{eqnarray}
From this equation, we obtain
\begin{eqnarray}
a(t) \propto \exp \left(\sqrt{\frac{{\Lambda}_{\mathrm{i}}}{3}} t \right)\,.
\label{eq:3.14}
\end{eqnarray}
Hence exponential inflation can be realized.
Thus, we see that the terms in $F(R)$ on the left-hand side of
Eq.~(\ref{eq:3.4}), i.e., the part of the braces $\{\}$, can be a
source of inflation, in addition to the terms in $f(R)$ on the right-hand side
of Eq.~(\ref{eq:3.4}). In fact,
if there do not exist any terms in $F(R)$, in which the theory is general
relativity, or the contribution of the terms in $F(R)$ to inflation is much
smaller than those in $f(R)$, Eq.~(\ref{eq:3.4}) is equivalent to
Eq.~(\ref{eq:2.41}).
In such a case, similarly to the consideration in Sec.~II\ C,
substituting $a(t) \propto t^{\tilde{p}}$, where $\tilde{p}$ is a 
positive constant, the approximate expressions of 
$f_{\mathrm{NO}}^{\prime}(R)$, 
$f_{\mathrm{NO}}^{\prime\prime}(R)$ and 
$f_{\mathrm{NO}}^{\prime\prime\prime}(R)$ derived from Eq.~(\ref{eq:3.12}) 
into Eq.~(\ref{eq:3.4}), we find $\tilde{p}= q+1$. 
Hence, if $q \gg 1$, $\tilde{p}$ becomes much larger than unity, so that
power-law inflation can be realized. 
Consequently, in the present model there exist two sources of inflation,
one from the modified part of gravity, $F(R)$, and the other from
the non-minimal gravitational coupling of
the electromagnetic field, $f(R)$. 
We here note that even if the value of ${\Lambda}_{\mathrm{i}}$ is so 
small that the modification of gravity cannot contribute to inflation, 
inflation can be realized due to the non-minimal gravitational coupling of
the electromagnetic field, namely, the change of the value of $f(R)$ 
in terms of $R$, and the generation of magnetic fields. 
This is an important feature of the present model. 

\subsection{Late-time cosmic acceleration}

Next, we consider the late-time acceleration of the universe. 
As shown above, at the early stage of the universe, at which the curvature is
very large, inflation can be realized due to the terms in $F(R)$ and/or
those in $f(R)$. As curvature becomes small, the contribution of these terms
to inflation becomes small, namely, the values of these terms in
Eq.~(\ref{eq:3.4}) become small, and then inflation ends. After inflation,
radiation becomes dominant, and subsequently matter becomes dominant.
When the energy density of matter becomes small and the value of curvature
becomes $R_0$, there appears the small effective cosmological constant
at the present time as seen in Eq.~(\ref{eq:3.7}). Hence, the current
cosmic acceleration can be realized.
It has been shown in Ref.~\cite{Nojiri:2007as}
that both inflation and the late-time acceleration of the universe
can be realized in modified gravitational theories described by
the action (\ref{eq:3.2}) with $F(R)$ in Eq.~(\ref{eq:3.5}) for
the case without the non-minimal gravitational
coupling of the electromagnetic field $f(R)$ in Eq.\ (\ref{eq:2.4}).
In this subsection, we confirm that also in this theory with
the non-minimal electromagnetic coupling $f(R)$,
the late-time acceleration of the universe can be realized.
(Incidentally, it has been shown in Ref.~\cite{Nojiri:2007bt} that
in this theory with a non-minimal coupling with the kinetic term of
a massless scalar field, the late-time acceleration of the universe
can be realized.)

In the limit $R \to R_0$, i.e., the present time,
because $R/M^2 -R_0/M^2 \ll 1$, we are able to use the following
approximate relations:
\begin{eqnarray}
F(R) \Eqn{\approx}
-M^2 \frac{c_3}{\left[ c_3 + c_4 \left( R_0/M^2 \right)^{2l+1} \right]^2}
\nonumber \\[2mm]
&& {}\times
\left\{ \left( \frac{R}{M^2} - \frac{R_0}{M^2} \right)^{2l+1} +
\left[ \frac{c_3 + c_4 \left( R_0/M^2 \right)^{2l+1}}{c_3} \right]
\left( \frac{R_0}{M^2} \right)^{2l+1}
\right\}\,,
\label{eq:3.15} 
\end{eqnarray}
and
\begin{eqnarray}
f_{\mathrm{NO}}(R) \Eqn{\approx}
\frac{c_5}{\left[ c_5 + c_6 \left( R_0/M^2 \right)^{2q+1} \right]^2}
\nonumber \\[2mm]
&& {}\times
\left\{ \left( \frac{R}{M^2} - \frac{R_0}{M^2} \right)^{2q+1} +
\left[ \frac{c_5 + c_6 \left( R_0/M^2 \right)^{2q+1}}{c_5} \right]
\left( \frac{R_0}{M^2} \right)^{2q+1}
\right\}\,.
\label{eq:3.16} 
\end{eqnarray}
From Eqs.~(\ref{eq:3.4}), (\ref{eq:3.15}), 
the approximate expressions of 
$F^{\prime}(R)$, 
$F^{\prime\prime}(R)$ and $F^{\prime\prime\prime}(R)$ derived from 
Eq.~(\ref{eq:3.15}), and 
the approximate expressions of 
$f_{\mathrm{NO}}^{\prime}(R)$, 
$f_{\mathrm{NO}}^{\prime\prime}(R)$ and 
$f_{\mathrm{NO}}^{\prime\prime\prime}(R)$ derived from 
Eq.~(\ref{eq:3.16}), 
we see that if $q>l$,
$f_{\mathrm{NO}}(R)$ becomes constant more rapidly than
$F(R)$ in the limit $R \to R_0$. As a result, the electrodynamics looks
as purely minimal theory at the current universe. For such a case, in the
limit $R \to R_0$, Eqs.~(\ref{eq:3.4}) are reduced to
\begin{eqnarray}
\dot{H} + H^2 = \frac{R_0}{3}\,.
\label{eq:3.17}
\end{eqnarray}
From this equation, we obtain
\begin{eqnarray}
a(t) \propto \exp \left(\sqrt{\frac{R_0}{3}} t \right)\,,
\label{eq:3.18}
\end{eqnarray}
so that
\begin{eqnarray}
\frac{\ddot{a}(t)}{a(t)} = \frac{R_0}{3} >0\,.
\label{eq:3.19}
\end{eqnarray}
Thus, the late-time acceleration of the universe can be realized.

Finally, we note the following point: 
In this model, 
even if the value of $R_0$ is so 
small that the modification of gravity cannot contribute to the late-time acceleration of the universe, the late-time acceleration can be realized 
due to due to the non-minimal gravitational coupling of 
the electromagnetic field and the generation of magnetic fields. 
This is also an important feature of the present model. 


\section{Classically equivalent form of non-minimal Maxwell-$F(R)$ gravity}

In this section, we consider classically equivalent form of non-minimal
Maxwell-$F(R)$ gravity.

The action (\ref{eq:3.1}) can be rewritten by using auxiliary fields.
Introducing two scalar fields $\zeta$ and $\xi$, we can rewrite
the action (\ref{eq:3.1}) to the following
form~\cite{matter-1, Nojiri:2007bt}:
\begin{eqnarray}
S \Eqn{=}
\int d^{4}x \sqrt{-g}
\left\{\,
\frac{1}{2\kappa^2} \left[ \zeta + F(\zeta) \right] + I(\zeta)
{\mathcal{L}}_{\mathrm{M}}
+ \xi \left( R - \zeta \right)
\, \right\}\,,
\label{eq:4.1} \\[2mm]
{\mathcal{L}}_{\mathrm{M}}
\Eqn{=} -\frac{1}{4} F_{\mu\nu}F^{\mu\nu}\,,
\label{eq:4.2}
\end{eqnarray}
where ${\mathcal{L}}_{\mathrm{M}}$ is the Lagrangian describing the ordinary
Maxwell theory. The form in Eq.~(\ref{eq:4.1}) is reduced to the original
form in Eq.~(\ref{eq:3.1}) by using the equation $\zeta=R$, which is derived
by taking variation of the action (\ref{eq:4.1}) with respect to
one auxiliary field $\xi$.
Moreover, taking variation of the form in Eq.~(\ref{eq:4.1}) with respect to
the other auxiliary field $\zeta$, we find
\begin{eqnarray}
\xi =
\frac{1}{2\kappa^2} \left[ 1 + F^{\prime}(\zeta) \right] +
I^{\prime}(\zeta) {\mathcal{L}}_{\mathrm{M}}\,,
\label{eq:4.3}
\end{eqnarray}
where the prime denotes differentiation with respect to $\zeta$.
Substituting Eq.~(\ref{eq:4.3}) into Eq.~(\ref{eq:4.1}) and eliminating
$\xi$ from Eq.~(\ref{eq:4.1}), we find
\begin{eqnarray}
S \Eqn{=}
\int d^{4}x \sqrt{-g}
\biggl\{\,
\frac{1}{2\kappa^2} \left[ 1 + F^{\prime}(\zeta) \right]R +
\left[ I(\zeta) + I^{\prime}(\zeta) \left( R - \zeta \right) \right]
{\mathcal{L}}_{\mathrm{M}} \nonumber \\[2mm]
&& \hspace{23mm}
{}+ \frac{1}{2\kappa^2}
\left[ F(\zeta) - F^{\prime}(\zeta) \zeta \right]
\, \biggr\}\,.
\label{eq:4.4}
\end{eqnarray}

We make the following conformal transformation of 
the action given by Eq.~(\ref{eq:4.4}): 
\begin{eqnarray}
g_{\mu \nu} \hspace{0.5mm} \rightarrow \hspace{0.5mm}
\hat{g}_{\mu \nu} = e^{\varphi} g_{\mu \nu}\,,
\label{eq:4.5}
\end{eqnarray}
with
\begin{eqnarray}
e^{\varphi} = 1 + F^{\prime}(\zeta)\,,
\label{eq:4.6}
\end{eqnarray}
where $\varphi$ is a scalar field. Here, the hat denotes quantities 
in a new conformal frame in which the term in the coupling between 
$F^{\prime}(\zeta)$ and the scalar curvature in the first term on the 
right-hand side of Eq.~(\ref{eq:4.4}) disappears. 
Consequently, the action in the new conformal frame is given by~\cite{F-M} 
\begin{eqnarray}
S_{\mathrm{N}} \Eqn{=}
\int d^{4}x \sqrt{-\hat{g}}
\boldmath{\Biggl[}
\frac{1}{2\kappa^2} \left( \hat{R} - \frac{3}{2} \hat{g}^{\mu\nu}
{\partial}_{\mu} \varphi {\partial}_{\nu} \varphi \right) \nonumber \\[2mm]
&& \hspace{0mm}
{}+ \left( e^{-2\varphi} \left\{ I\left[\zeta(\varphi) \right] -
I^{\prime}\left[\zeta(\varphi) \right] \zeta(\varphi) \right\}
+ e^{-\varphi} I^{\prime}\left[\zeta(\varphi) \right]
\left( \hat{R}+3\hat{\Box} \varphi - \frac{3}{2} \hat{g}^{\mu\nu}
{\partial}_{\mu} \varphi {\partial}_{\nu} \varphi \right)
\right) \hat{\mathcal{L}}_{\mathrm{M}} \nonumber \\[2mm]
&& \hspace{0mm}
{}+ \frac{1}{2\kappa^2} e^{-2\varphi}
\left\{ F\left[\zeta(\varphi) \right]
- \left( e^{\varphi} - 1 \right) \zeta(\varphi) \right\}
\boldmath{\Biggr]}\,,
\label{eq:4.7}
\end{eqnarray}
where
\begin{eqnarray}
\hat{\Box} \varphi = \frac{1}{\sqrt{-\hat{g}}} {\partial}_{\mu}
\left( \sqrt{-\hat{g}} \hat{g}^{\mu\nu} {\partial}_{\nu} \varphi
\right)\,,
\label{eq:4.8}
\end{eqnarray}
and $\hat{g}$ is the determinant of $\hat{g}^{\mu\nu}$. In deriving
Eq.~(\ref{eq:4.7}), we have used Eq.~(\ref{eq:4.6}).
Moreover, $\zeta(\varphi)$ in Eq.~(\ref{eq:4.7}) is obtained by solving
Eq.~(\ref{eq:4.6}) with respect to $\zeta$ as $\zeta=\zeta(\varphi)$.
Hence, the action in the new conformal frame (\ref{eq:4.7}) includes 
the Brans-Dicke type scalar field $\varphi$~\cite{B-D}. 
From the term proportional to $\hat{\mathcal{L}}_{\mathrm{M}}$ 
on the right-hand side of Eq.~(\ref{eq:4.7}), 
we see that the form of the Lagrangian in terms of 
the electromagnetic field in Eq.~(\ref{eq:4.7}) is close to 
that of the electromagnetic field with the coupling to the dilaton, which 
has been explained in Sec.\ II~C. In other words, 
the Lagrangian of non-minimal Maxwell-$F(R)$ gravity is 
qualitatively similar to Lagrangian describing 
dilaton electromagnetism (except for the term in the coupling between 
the scalar curvature and the electromagnetic field). 
As explained in Refs.~\cite{equivalence, Faraoni:2006fx}, however,
this fact does not mean the physical equivalence between them.

\section{Conclusion}

In the present paper, we have considered inflation and the late-time
acceleration in the expansion of the universe in non-minimal electromagnetism,
in which the electromagnetic field couples to the scalar curvature function.
As a result, we have shown that power-law inflation can be realized due to
the non-minimal gravitational coupling of the electromagnetic field,
and that large-scale magnetic fields can be generated due to the breaking of
the conformal invariance of the electromagnetic field through its non-minimal
gravitational coupling.
Furthermore, we have demonstrated that both inflation and the late-time
acceleration of the universe can be realized in a modified Maxwell-$F(R)$
gravity proposed in Ref.~\cite{Nojiri:2007as}
which is consistent with solar-system tests and cosmological bounds
and free of instabilities.
We have also considered classically equivalent form of non-minimal
Maxwell-$F(R)$ gravity.

Finally, we make a remark about the observational deviation of a non-minimal
electromagnetic theory from the ordinary Maxwell theory.
It follows from the fourth equation in Eq.~(\ref{eq:2.9}) that
in exponential inflation the scalar curvature is proportional to
the square of the Hubble parameter. Moreover, it is known that
the root-mean-square (rms) amplitude of curvature perturbations is also
proportional to the square of the Hubble parameter.
In a non-minimal electromagnetic theory, because magnetic fields
couple to the scalar curvature,
there can exist the cross correlations between magnetic fields and
curvature perturbations through the Hubble parameter. Hence,
if the primordial large-scale magnetic fields are
detected~\cite{Caprini:2003vc, Kahniashvili:2006zs}
by future experiments such as PLANCK \cite{Planck}, SPIDERS
(post-PLANCK) \cite{SPIDERS} and Inflation Probe (CMBPol mission) in the Beyond Einstein program of NASA \cite{CMBPol} on the anisotropy of CMB radiation,
and if there exist (do not exist) the cross correlations between
the primordial large-scale magnetic fields and curvature perturbations,
it is observationally suggested that
at the inflationary stage there should exist a non-minimal gravitational
coupling of the electromagnetic field
(the strength and/or the form of non-minimal coupling of the
electromagnetic field may be observationally restricted).


\section*{Acknowledgements}
We are grateful to M.~Sasaki and S.~Nojiri for very helpful discussion of
related problems.
The work of K.B. was supported in part by the
open research center project at Kinki University and that by S.D.O. was
supported in part by MEC (Spain) projects FIS2006-02842 and
PIE2007-50/023.

\appendix*
\section{Asymptotic freedom versus non-minimal coupling}

It is very interesting that one can generalize the discussion of this work
for interacting theories: scalar/spinor electrodynamics and non-Abelian
gauge theory. As a simple example, let us consider the $SU(2)$ gauge theory
with the Lagrangian:
$\mathcal{L}=-\left(1/4 \right)
G^{a}_{\mu\nu}G^{a\mu\nu}$, where $G^{a}_{\mu\nu}$ is the $SU(2)$ field
strength.
The effective renormalization-group improved Lagrangian for such a theory in
matter sector has been found in Ref.~\cite{Elizalde:1996am} for a de Sitter
background as
\begin{eqnarray}
{\mathcal{L}}_{SU(2)}
=-\frac{1}{4} \frac{\tilde{g}^2}{\tilde{g}^2(\tilde{t})}
G^{a}_{\mu\nu}G^{a\mu\nu}\,,
\label{eq:A.1}
\end{eqnarray}
with
\begin{eqnarray}
\tilde{g}^2(\tilde{t})
=\frac{\tilde{g}^2}{1+11\tilde{g}^2\tilde{t}/\left(12 \pi^2\right)}\,,
\label{eq:A.2}
\end{eqnarray}
where $\tilde{g}(\tilde{t})$ is the running $SU(2)$ gauge coupling constant,
$\tilde{g}$ is the value of $\tilde{g}(\tilde{t})$ in the case $\tilde{t}=0$,
and $\tilde{t}$ is a renormalization-group parameter.  Note that the running
gauge coupling constant typically shows asymptotically free behavior:
it goes to zero at very high energy.
For the covariantly constant gauge background with
$G^{a}_{\mu\nu}G^{a\mu\nu}/2 = \tilde{H}^2$, where $\tilde{H}$ corresponds
to the magnetic field in the $SU(2)$ gauge theory, it has been proposed in
Ref.~\cite{Elizalde:1996am} that $\tilde{t}$ is given by
\begin{eqnarray}
\tilde{t}
= \frac{1}{2} \ln \frac{R/4 + \tilde{g} \tilde{H}}{\mu^2} \,,
\label{eq:A.3}
\end{eqnarray}
where $\mu$ is a mass parameter.

It is clear that with the decrease of the energy scale
(namely, as the universe expands),
$\tilde{t}$ is decreasing, as $\tilde{t}$ is very large at the early
universe.
Taking into account the results of this work, one can try to relate the
asymptotic freedom in a non-Abelian gauge theory with non-minimal
Maxwell-modified gravity.
In this way, using the proposal of Eq.~(\ref{eq:2.42})
in Sec.\ II for non-minimal $f(R)$ in front of $G^{a}_{\mu\nu}G^{a\mu\nu}$,
one gets
\begin{eqnarray}
\frac{c_1 \left(R/m^2 \right)^n}{c_2 \left(R/m^2 \right)^n + 1}
=\frac{11\tilde{g}^2}{12 \pi^2} \tilde{t}\,.
\label{eq:A.4}
\end{eqnarray}
Hence, according to this assumption, at a very large curvature
($R/m^2 \gg 1$),
$\tilde{t} \approx \left[12 \pi^2/\left( 11\tilde{g}^2 \right) \right]
\left(c_1/c_2\right)$,
while at the current universe ($R \to 0$), $\tilde{t} \to 0$.
Thus, asymptotic freedom induces the appearance of the non-minimal
gravitational gauge coupling in (non-) Abelian gauge theories at high
energy.

Generally speaking, such a scenario is universal and it works not only for
asymptotically free theories. For instance, for scalar QED
one can easily write the renormalization-group improved effective
Lagrangian in curved spacetime.
In the matter sector (zero scalar field background) it has
qualitatively the same form as Eq.~(\ref{eq:A.1}),
the only sign of $\tilde{t}$ is different
in the expression for the running gauge coupling constant. As a result,
such an effective Lagrangian again induces the non-minimal gravitational
coupling of the electromagnetic sector.


\end{document}